\documentclass[onecolumn,nofootinbib,12pt]{article}
\usepackage{jheppub}
\usepackage{ifpdf}
\usepackage{graphicx}
\usepackage{amsfonts}
\usepackage{amsmath}
\usepackage{amssymb}
\usepackage{epsfig}
\usepackage{pdfpages}
\usepackage{graphicx,epstopdf}
\usepackage{caption}
\usepackage[position=b]{subcaption}
\usepackage[makeroom]{cancel}
\usepackage{hyperref}
\usepackage{array}

\hypersetup{pdftex,colorlinks=true,allcolors=blue}
\def\pa{\partial}

\newcommand{\br}{\biggr}
\newcommand{\bl}{\biggl}

\begin{document}

\title{Phase transitions of neutral planar hairy AdS black holes}

\author[a]{Andres Anabalon,}
\author[b]{Dumitru Astefanesei,}
\author[b,e]{David Choque,}
\author[d]{Jose D. Edelstein}
\affiliation[a]{Departamento de Ciencias, Facultad de Artes Liberales,Universidad Adolfo Ib\'{a}\~{n}ez, Av. Padre Hurtado 750, Vi\~{n}a del Mar, Chile.}
\affiliation[b]{Instituto de F\'\i sica, Pontificia Universidad Cat\'olica de Valpara\'\i so, Casilla 4059, Valpara\'{\i}so, Chile.}
\affiliation[c]{Universidad T\'{e}cnica Federico Santa Mar\'{\i}a, Av. Espa\~{n}a 1680, Valpara\'{\i}so, Chile.}
\affiliation[d]{Departamento de F\'\i sica de Part\'\i culas $\&$ Instituto Galego de F\'\i sica de Altas Enerx\'\i as (IGFAE), Universidad de Santiago de Compostela, E-15782 Santiago de Compostela, Spain}
\affiliation[e]{Universidad Nacional de San Antonio Abad del Cusco, Av. La Cultura 733, Cusco, Per\'u.}

\date{\today}


\emailAdd{andres.anabalon@uai.cl} \emailAdd{dumitru.astefanesei@pucv.cl} %
\emailAdd{brst1010123@gmail.com} \emailAdd{jose.edelstein@usc.es}

\abstract{We investigate the phase diagram of a general class of $4$-dimensional exact regular hairy planar black holes. For some particular values of the parameters in the moduli potential, these solutions can be embedded in $\omega$-deformed $\mathcal{N}=8$ gauged supergravity. We construct the hairy soliton that is the ground state of the theory and show that there exist first order phase transitions.}

\maketitle

\section{Introduction}

The AdS/CFT correspondence \cite{Maldacena:1997re} enables us to understand phases of strongly coupled gauge theories as well as their phase transitions from a dual perspective. Famously enough, Hawking and Page showed \cite{Hawking} that there exists a phase transition between spherical AdS (Schwarzschild) black hole and global AdS spacetime. Above the critical temperature, $T_c$, the black hole becomes thermodynamically stable while below $T_c$ it is unstable, global AdS being energetically favorable. Within the AdS/CFT duality this corresponds to a confinement/deconfinement phase transition in the dual quantum field theory \cite{Witten:1998zw}. On the Euclidean section there are two scales: the temperature (related to the periodicity of the Euclidean time, $\beta$) and the radius of the 3-sphere, $\beta^\prime$. The black hole represents a thermal state in the field theory and so, due to conformal invariance, only the ratio $\beta/\beta^\prime$ is relevant for the phase transition of the dual gauge theory on $S^3 \times S^1$.

The phase transition is sensitive to the topology of spacetime, both on the gravity side as well as on the gauge theory side. For AdS black holes with planar horizon geometry, for instance, there exists no Hawking-Page transition with respect to thermal AdS: the planar black hole phase is dominant for any non-zero temperature. This can be understood from the fact that planar geometry can be obtained from spherical geometry in the $\beta^\prime \rightarrow \infty $ limit, thereby $\beta/\beta^\prime \rightarrow 0$. Correspondingly, the dual thermal gauge theory, which is defined on $R^3 \times S^1$, does not display any phase transition. The gauge theory is always in the deconfined phase. However, when (at least) one of the spatial directions is compactified asymptotically on a circle, there exists a negative Casimir energy of the non-supersymmetric field theory that `lives' on such  non-trivial topology.

The main question then is what is the corresponding bulk geometry? Using a double analytic continuation --- both in the time and compactified angular directions --- of the planar black hole, Horowitz and Myers showed \cite{Horowitz:1998ha} that, indeed, there exists a solution with a lower energy than AdS itself, which was dubbed \textit{the AdS soliton}. This solution can be understood as the ground state of the theory. The proof of the uniqueness of the AdS soliton was originally sketched in \cite{Horowitz:1998ha} and finally demonstrated in \cite{Galloway:2001uv, Galloway:2002ai} (see also \cite{Woolgar:2016axs}). The periodicity in the Euclidean time direction is unconstrained, thereby the AdS soliton exists at all temperatures. This fits very nicely with the expectation that a non-supersymmetric gauge theory could be described within the AdS/CFT duality by compactifying one direction and imposing anti-periodic boundary conditions for the fermions around the circle \cite{Witten:1998zw}. Interestingly enough, there exists a Hawking-Page phase transition between the planar black hole and the thermal AdS soliton \cite{Surya:2001vj}.

When we think of the embedding of these solutions in supergravity, we need to consider the possibility of turning on the remaining massless fields; in particular, the real scalars arising from the compact manifold used in the dimensional reduction. This triggers the interest in the role of scalar hair in the framework of phase transitions. The thermodynamics of hairy black holes in AdS, e.g. \cite{Hertog:2004bb, Anabalon:2012ta, Acena:2012mr, Anabalon:2013qua, Acena:2013jya, Feng:2013tza, Fan:2015tua}, has been recently studied in a series of papers \cite{Lu:2013ura, Anabalon:2013sra, Anabalon:2015ija, Astefanesei:2018vga, Astefanesei:2019mds}.\footnote{In a recent paper \cite{Astefanesei:2019qsg}, it was proven that similar, but spherical, black hole solutions in flat space are thermodynamically and dynamically stable.} The model we are interested in contains a scalar potential depending on two parameters \cite{Anabalon:2012ta}, whose embedding in so-called $\omega$-deformed $\mathcal{N}=8$ gauged supergravity \cite{DallAgata:2012mfj} (see \cite{Trigiante:2016mnt} for a nice review) was performed in \cite{Anabalon:2013eaa, Tarrio:2013qga}. The Hawking-Page phase transition also holds for these spherical hairy black holes \cite{Anabalon:2015ija}, while for the planar case the arguments discussed above apply and there is no first order phase transition.

In this paper we shall explore the phases of four-dimensional neutral (hairy) planar black holes and solitons in AdS spacetime with a spatial direction compactified on a circle. This analysis is important in the context of the AdS/CFT duality, where the phase diagram of the bulk gravity theory provides important qualitative information about the features of the dual theory at finite temperature. We shall consider a general Einstein-dilaton theory whose dynamics is governed by the action
\begin{equation}
I[g_{\mu\nu},\phi] = \int_{\mathcal{M}} d^{4}x \sqrt{-g} \left[ \frac{R}{2\kappa} - \frac12 (\partial\phi)^2 - V(\phi) \right] + \frac{1}{\kappa} \int_{\partial\mathcal{M}} d^{3}x \sqrt{-h}\, K ~,
\label{action}
\end{equation}
with the following potential \cite{Anabalon:2013sra}:
\begin{eqnarray}
V(\phi) &=& \frac{\Lambda}{6} \frac{\nu^{2}-4}{\nu^{2}} \left[ \frac{\nu-1}{\nu+2} e^{-\ell_\nu^+ \phi} + \frac{\nu+1}{\nu-2} e^{\ell_\nu^- \phi} + 4 \frac{\nu^{2}-1}{\nu^{2}-4} e^{-\ell_\nu \phi} \right]  \notag \\ [0.3em]
& & + \frac{\alpha}{\nu^{2}} \left[ \frac{\nu-1}{\nu+2} \sinh{\ell_\nu^+ \phi} - \frac{\nu+1}{\nu-2} \sinh{\ell_\nu^- \phi} + 4 \frac{\nu^{2}-1}{\nu^{2}-4} \sinh{\ell_\nu \phi} \right] ~,
\label{potential}
\end{eqnarray}
depending on three real free parameters $\Lambda$, $\alpha$ and $\nu$, where $\ell_\nu^{-1} = \sqrt{(\nu^2-1)/2}$ and $\ell_\nu^\pm = (\nu \pm 1)\,\ell_\nu$. Written in this way, the potential has two parts: one proportional to $\Lambda$ and the other one proportional to $\alpha$. Asymptotically the scalar field vanishes (so it does the term proportional
to $\alpha$), the only contribution to the potential becoming $V(\phi=0)=\Lambda$; therefore $\Lambda$ can be interpreted as the cosmological constant. When $\alpha=0$, this potential corresponds to a truncation of $\mathcal{N}=2$ gauged supergravity \cite{Faedo:2015jqa}, and when $\alpha^2 < -\Lambda/3$ it is a truncation of $\mathcal{N}=2$ supergravity with an electromagnetic gauging \cite{Anabalon:2017yhv} (see, also, \cite{Gallerati:2019mzs}). More interestingly, for some specific values of the potential parameters, we can explicitly obtain the embedding in $\omega$-deformed $\mathcal{N}=8$ gauged supergravity and exact hairy black hole solutions \cite{Anabalon:2013eaa}. The additional parameter $\nu$ is referred to as the \textit{hair parameter} in the sense that when $\nu=\pm 1$ the usual gravity theory with negative cosmological constant is obtained $V(\phi, \nu=\pm 1) = \Lambda$. Notice that the action is invariant under the combined transformation $\phi \to -\phi$ and $\alpha \to - \alpha + \frac{\Lambda}{3} (\nu^2 - 4)$. 

We shall use as a reference the exact planar AdS black hole solution of the theory given by (\ref{action}) and (\ref{potential}), which correspond to mixed boundary conditions for the scalar field \cite{Anabalon:2013eaa,
Acena:2013jya}, for the sake of understanding the phase diagram of gravitational solutions with the same asymptotics. Indeed, in the same class of boundary conditions there exists also a hairy AdS soliton for which the integration constant is fixed by the periodicity of the angular coordinate such that the conical singularity is eliminated. Since the latter has a compact spatial direction, there is a negative Casimir energy contribution similar to the one in \cite{Horowitz:1998ha}. The contribution of the scalar to the mass of hairy black holes and hairy solitons, and their relation to the boundary conditions of the scalar field, were discussed in  \cite{Hertog:2004dr, Henneaux:2006hk, Dolan:2015dha, Ponglertsakul:2016fxj, Harms:2016pow, Naderi:2019jhn, Ong:2019glf, Astefanesei:2019pfq} (see also \cite{ Astefanesei:2008wz, Astefanesei:2010dk, Cadoni:2011yj, Brihaye:2012ww, Brihaye:2013tra, Kleihaus:2013tba} for various applications in AdS and flat spacetimes), though they typically rely in numerical methods. In  \cite{Anabalon:2014fla}, by making use of the Hamiltonian formalism, it was shown that when the conformal symmetry is broken at the boundary there is an extra contribution to the mass of hairy black holes that makes the first law of thermodynamics well posed, explaining some of the previous results. 

For a particular scalar potential (when a parameter of the scalar potential proposed in \cite{Anabalon:2013sra} is fixed), the exact hairy AdS soliton solutions were constructed in \cite{Anabalon:2016izw}. We generalize this work by including all three independent parameters of the potential, constructing the most general exact hairy AdS soliton solution, and making a detailed analysis of the first order phase transitions arising for different values in the space of parameters. 

The structure of the paper is as follows. In Section \ref{section2}, after a short review  of the AdS soliton in pure gravity, we present the hairy black hole solutions and construct the corresponding hairy AdS soliton. We compute their mass and free energy by using the counterterm method of Balasubramanian and Kraus supplemented with extra counterterms for the scalar field as was proposed in \cite{Anabalon:2015xvl}. We also obtain the regularized  stress tensor of the dual field theory for the hairy black hole and soliton in the form of a fluid \cite{Myers:1999psa}. In Section 3, we present a detailed analysis of the impact of hair on the phase transitions and show that there exist first order phase transitions between the hairy black hole and the AdS soliton with the same boundary conditions. In Section 4, we obtain an analytic holographic c-function and show that in the limit when the hair is turned off, $\nu=\pm 1$, the flow is trivial. Finally, in Section 5 we summarize our results.

\section{Exact hairy AdS soliton solutions}
\label{section2}

Let us briefly review the construction of the AdS soliton \cite{Horowitz:1998ha} and its main properties, which will be necessary as a basis to construct the class of exact hairy AdS solitons we are interested in.

\subsection{The AdS soliton}

We start with the usual Einstein-Hilbert action including the Gibbons-Hawking boundary term: 
\begin{equation}
I[g_{\mu\nu}] = \frac{1}{\kappa} \int_{\mathcal{M}} d^{4}x \sqrt{-g} \left[ \frac{R}{2} - \Lambda \right] + \frac{1}{\kappa} \int_{\partial\mathcal{M}} d^{3}x \sqrt{-h}\, K ~,
\label{actionSchw}
\end{equation}
where $\Lambda = -3/l^2$ is the cosmological constant, $l$ is the AdS radius, $K$ is the boundary extrinsic curvature, and $h_{ij}$ is the induced metric at the boundary. It is well known that this theory possesses black hole solutions with planar horizon topology \cite{Horowitz:1998ha},
\begin{equation}
ds^{2} = - \left( - \frac{\mu_{b}}{r} + \frac{r^{2}}{l^{2}} \right) dt^{2} + \left( - \frac{\mu_{b}}{r} + \frac{r^{2}}{l^{2}} \right)^{-1} dr^{2} + \frac{r^{2}}{l^{2}} \left( dy^{2} + dz^{2} \right) ~,
\label{bh}
\end{equation}
where $\mu_{b}$ is the only integration constant, the mass parameter, and we are considering the following compactified coordinates: $0 \leq y \leq L_{b}$ y $0 \leq z \leq L$. Interestingly, Horowitz and Myers \cite{Horowitz:1998ha} used a double analytic continuation (in time and one of the compactified directions, $t \rightarrow i\theta$ and $y \rightarrow i\tau $) to construct a new solution:
\begin{equation}
ds^{2} = - \frac{r^{2}}{l^{2}} d\tau^{2} + \left( - \frac{\mu_{s}}{r} + \frac{r^{2}}{l^{2}} \right)^{-1} dr^{2} + \left( - \frac{\mu_{s}}{r} + \frac{r^{2}}{l^{2}} \right) d\theta^{2} + \frac{r^{2}}{l^{2}} dz^{2} ~,
\label{Schwsol}
\end{equation}
where we denoted the mass parameter by $\mu_s$ (and the periodicity of the new angular coordinate $\theta$ will be called $L_s$) in order to neatly distinguish between the AdS soliton and the planar black hole solutions. The solution (\ref{Schwsol}) should be regular and have Lorentzian signature. The first condition is fulfilled by eliminating a conical singularity in the $(r, \theta)$ plane, which leads us to
\begin{equation}
L_{s} = \frac{4\pi\sqrt{g_{\theta\theta}\, g_{rr}}}{(g_{\theta\theta})^\prime}\biggr{\vert}_{r=r_{s}} = \frac{4\pi l^{2}}{3r_{s}} ~,
\label{periosol}
\end{equation}
where $r_s$ is the radius of the AdS soliton, $r_s^3 = \mu_s\,l^2$. The second condition further restricts the radial coordinate, $r \geq r_{s}$. One can readily compute the Euclidean action
\begin{equation}
I^{E}_{s} = - \frac{LL_{s}\beta_{s}\mu_{s}}{2\kappa\,l^{2}} ~,
\label{Schwactsol}
\end{equation}
where $\beta_s$ is nothing but the periodicity of the Euclidean time $\tau \rightarrow i\tau_E$, such that the AdS soliton is at the finite temperature, $T=\beta_s^{-1}$. The mass can be obtained as usual by using its thermodynamical relation with the free energy,
\begin{equation}
\mathcal{F} = I^{E}_{s}/\beta_{s} = M \quad \Rightarrow \quad M = - \frac{LL_{s}\mu_{s}}{2\kappa\, l^{2}} ~.
\end{equation}
Since its mass is negative --- thereby, smaller than the free energy of AdS itself ---, it was proposed that the AdS soliton is the ground state of the theory \cite{Horowitz:1998ha}. A negative mass can be interpreted as the Casimir energy that is generated in the dual filed theory due to fermions being antiperiodic along the spatial compact coordinate \cite{Witten:1998zw}.

\subsection{The hairy AdS soliton}

We want to study hairy AdS solitons whenever the scalar field sector in (\ref{action}) and (\ref{potential}) is turned on. In order to obtain an exact planar black hole, which we will need as a seed to construct the soliton, we use the following ansatz for the metric \cite{Anabalon:2012ta}:
\begin{equation}
ds^{2} = \Omega_b(x) \left[ - f(x) dt^{2} + \frac{\eta_b^{2}dx^{2}}{f(x)} + \frac{dy^{2}}{l^{2}} + \frac{dz^{2}}{l^{2}}\right] ~,
\label{Ansatzbh1}
\end{equation}
where $x$ is a more convenient radial variable, $x \leq 1$; one can easily check that the (conformal) boundary is at $x=1$ and the metric is asymptotically AdS. The equations of motion can be readily integrated to obtain the conformal factor \cite{Anabalon:2012ta, Acena:2013jya, Anabalon:2013sra, Anabalon:2013qua, Acena:2012mr},
\begin{equation}
\Omega_b(x) = \frac{\nu^{2}x^{\nu-1}}{\eta_b^{2}(x^{\nu}-1)^{2}} ~.
\label{omega}
\end{equation}
We need to include the integration constant $\eta_b$, related to the mass of the black hole; since the \textit{hair} is secondary there is no integration constant associated to it. The dilaton adopts a quite simple form in terms
of the radial variable $x$,
\begin{equation}
\phi(x)= \ell_{\nu}^{-1} \ln{x} ~,
\end{equation}
and the metric function reads:
\begin{equation}
f(x) = \frac{1}{l^{2}} + \alpha \left[ \frac{1}{\nu^{2}-4} - \frac{x^{2}}{\nu^{2}} \left( 1 + \frac{x^{-\nu}}{\nu-2}-\frac{x^{\nu}}{\nu+2} \right) \right] ~.
\label{f}
\end{equation}
We have now all the ingredients to construct the hairy AdS soliton. The procedure is similar to that of Horowitz and Myers; we use the same double analytical continuation and obtain the following metric:
\begin{equation}
ds^{2} = \Omega_s(x) \left[ -\frac{d\tau^2}{l^{2}} + \frac{\eta_s^{2}dx^{2}}{f(x)} + f(x) d\theta^{2} + \frac{dz^{2}}{l^{2}} \right] ~,
\label{anzatsol1}
\end{equation}
where the conformal factor can be obtained from (\ref{omega}) by replacing the black hole's mass parameter $\eta_b$ with $\eta_s$, which characterizes the hairy AdS soliton,
\begin{equation}
\Omega_s(x) = \frac{\nu^{2}x^{\nu-1}}{\eta_s^{2}(x^{\nu}-1)^{2}} ~.
\label{omegasol}
\end{equation}
The metric (\ref{anzatsol1}) has a conical singularity in the $(x,\theta)$ plane; thereby, to get a regular solution, we have to impose the following periodicity on the angular coordinate $\theta$:
\begin{equation}
L _{s} = \frac{4\pi\eta_s}{f^\prime(x)}\biggr{\vert}_{x=x_h} = \frac{4\pi\eta_s^{3}\,\Omega_s(x_h)}{\alpha} ~,
\label{periosoliton}
\end{equation}
where $x_h$ is the real root of $f(x_h)=0$.

\subsection{The Euclidean action of the hairy black hole}

Next we want to describe in detail the evaluation of the gravitational action and quasilocal stress tensor for both the hairy black hole and hairy AdS soliton. We need to regularize the action, which for pure gravity is performed \cite{Balasubramanian:1999re} by supplementing it with the counterterm $I_{\mathrm{count}}$:
\begin{equation}
I[g_{\mu\nu}] = I_{\mathrm{bulk}} + I_{\mathrm{surf}} + I_{\mathrm{count}} ~,
\end{equation}
where $I_{\mathrm{bulk}}$ and $I_{\mathrm{surf}}$ are the two terms in (\ref{action}), and
\begin{equation}
I_{\mathrm{count}} = - \frac{1}{\kappa}\int_{\partial\mathcal{M}} d^{3}x \sqrt{-h} \left( \frac{2}{l} + \frac{l}{2} \mathcal{R} \right) ~,
\label{Icount}
\end{equation}
$\mathcal{R}$ being the Ricci scalar of the boundary metric $h_{ab}$, which in this case vanishes. From now on we use the following notation: $x = 1 - \epsilon$ ($\epsilon \to 0^+$) and $x_h$ are the boundary and horizon locations, and $\beta_{b}$ is the periodicity of the Euclidean time that is related to the temperature of the hairy black hole, $\beta_{b} = T^{-1}$,
\begin{equation}
T = \frac{f^{\prime}(x)}{4\pi\eta_b}\biggr{\vert}_{x=x_h} = \frac{\alpha}{4\pi\eta_b^{3}\Omega_b(x_h)} ~.
\label{temperature}
\end{equation}
The Euclidean bulk action can be easily evaluated, 
\begin{equation}
I_{\mathrm{bulk}}^{E} = \frac{LL_b\beta_b}{2\kappa l^{2}\eta_b} \frac{d(f\Omega_b)}{dx}\biggr{\vert}_{x_h}^{1 - \epsilon} ~.
\end{equation}
To obtain the Gibbons-Hawking surface term, we choose a foliation $x=$ constant, which results in the induced metric:
\begin{equation}
ds^{2} = h_{ab} dx^{a}dx^{b} = \Omega_b(x) \left[ - f(x) dt^{2} + \frac{dy^{2}}{l^{2}} + \frac{dz^{2}}{l^{2}}\right] ~,
\label{xcte}
\end{equation}
while the normal to the boundary and its extrinsic curvature read 
\begin{equation}
n_{a} = \frac{\delta_{a}^{\ x}}{\sqrt{g^{xx}}} ~, \qquad K_{ab} = \frac{\sqrt{g^{xx}}}{2}\partial_{x}h_{ab} ~, \qquad K \sqrt{-h} = n^{a} \partial_{a} \sqrt{-h} ~.
\label{normalK}
\end{equation}
All in all, the contribution of the Gibbons-Hawking surface term to the Euclidean action can be written as follows:
\begin{equation}
I_{\mathrm{surf}}^{E} = - \frac{LL_{b}\beta_{b}}{2\kappa l^{2}\eta_b} \left[ (f\Omega_b)^\prime + 2 f \Omega_b^\prime \right]\bigg|_{1 - \epsilon} ~.
\end{equation}
When the scalar field is introduced, the action should also be supplemented with a counterterm that depends on it in order to get rid of all divergences \cite{Anabalon:2015xvl}:
\begin{equation}
I_{\phi} = \frac{1}{\kappa l} \int_{\partial\mathcal{M}} d^{3}x \sqrt{-h} \left( \frac{1}{2} \phi^{2} - \frac{\ell_{\nu}}{6} \phi^{3} \right) ~;
\label{scalarct}
\end{equation}
the Euclidean action, when evaluated on the scalar field configuration, reads:
\begin{equation}
I_{\phi}^{E} = - \frac{L L_{b}\beta_{b}}{3 \kappa l^{4}\eta_b^{3}} (1 - \nu^2) \left[ 1 + \frac{3}{4\epsilon} \right] ~.
\label{Iphi2}
\end{equation}
The sum of the first three terms in the action is 
\begin{equation}
I_{\mathrm{bulk}}^{E} + I_{\mathrm{surf}}^{E} + I_{\mathrm{count}}^{E} = - \frac{\mathcal{A}}{4G_{N}} + \frac{L L_{b}\beta_{b}}{3 \kappa l^{4}\eta_b^{3}} (1 - \nu^2) \left[ 1 + \frac{3}{4\epsilon} + \frac{\alpha l^{2}}{1 - \nu^2} \right] ~,
\end{equation}
where $\mathcal{A}=LL_{b}\Omega_b(x_h)/l^{2}$ is the area of the horizon; recall that $\kappa = 8\pi G_N$. It is clear now that the gravitational counterterm \cite{Balasubramanian:1999re} is not sufficient to cancel the divergence in the action because there is still a term proportional to $\epsilon^{-1}$. However, when the counterterm (\ref{Iphi2}) is added, the
divergence cancels out and we obtain a finite action:
\begin{equation}
I^{E}_{\mathrm{hairy\,BH}} = \beta_{b} \left( - \frac{\mathcal{A}T}{4G_{N}} + \frac{LL_{b}}{\kappa l^{2}}\frac{\alpha}{3\eta_b^{3}} \right) = - \frac{LL_{b}\alpha\beta_{b}}{6\kappa l^{2}\eta_b^{3}} = - \frac{2\pi T L L_b}{3 \kappa l^2} \beta_b \Omega_b(x_h) ~.
\label{Ibh}
\end{equation}
We can now compute the free energy, $\mathcal{F} = \beta_{b}^{-1} I^{E}_{\mathrm{hairy\,BH}}$, and by means of $\mathcal{F} = E - T S$ the hairy black hole energy can be computed,
\begin{equation}
E = \frac{LL_{b}\,\mu_{b}}{\kappa l^{2}} ~, \qquad \mu_{b} = \frac{\alpha}{3\eta_b^{3}} ~.
\label{en1}
\end{equation}
%

\subsection{The Brown-York stress tensor}

The regularized quasilocal stress tensor of Brown and York \cite{Brown:1992br} can be obtained by using the gravitational action supplemented with the counterterms:
\begin{equation}
\tau_{ab}=-\frac{1}{\kappa} \left( K_{ab} - h_{ab} K + \frac{2}{l} h_{ab} - l G_{ab} \right) - \frac{h_{ab}}{\kappa l} \left( \frac{\phi^{2}}{2}-\frac{\ell_{\nu}}{6}\phi^{3} \right) ~,
\label{stresstensor}
\end{equation}
We evaluate this expression in the hairy black hole configuration by using the expansion around the conformal boundary, $\epsilon \to 0^+$, and the components read
\begin{equation}
\tau_{tt} \simeq \frac{\alpha\epsilon}{3\kappa l\eta_b^{2}} ~, \qquad
\tau_{yy} \simeq \frac{\alpha\epsilon}{6\kappa l\eta_b^{2}} ~, \qquad
\tau_{zz} \simeq \frac{\alpha\epsilon}{6\kappa l\eta_b^{2}} ~,
\label{BYcomponents}
\end{equation}
to leading order in $\epsilon$. The induced boundary metric becomes
\begin{equation}
ds^{2} = \frac{R^{2}}{l^{2}} \left[ - dt^{2} + dy^{2} + dz^{2} \right] ~,
\label{Bblack}
\end{equation}
where the conformal factor, $\Omega_b(1-\epsilon) = R^2 + \mathcal{O}(\epsilon)$, where
\begin{equation}
R^{2} = \frac{1}{\eta_b^{2} \epsilon^{2}} ~.
\label{confactor}
\end{equation}
The boundary metric (\ref{Bblack}) is related, up to the conformal factor, to the dual metric
\begin{equation}
ds_{\mathrm{dual}}^{2} = \frac{l^{2}}{R^{2}} ds^{2} = \gamma_{ab} dx^{a} dx^{b} = - dt^{2} + dy^{2} + dz^{2} ~,
\label{dualmetric}
\end{equation}
thereby the stress tensor of the dual field theory is also rescaled \cite{Myers:1999psa}
\begin{equation}
\langle\tau^{\mathrm{dual}}_{ab}\rangle = \lim_{R \to \infty} \frac{R}{l} \tau_{ab} = \lim_{\epsilon \to 0} \frac{1}{l \eta_b \epsilon} \tau_{ab} = \frac{\mu_b}{2\kappa l^{2}} \left[ 3 \delta_{a}^{\ 0} \delta_{b}^{\ 0} + \gamma_{ab} \right] ~.
\label{taudual}
\end{equation}
It is covariantly conserved and traceless, as expected for a scalar field with mixed boundary conditions that preserve conformal symmetry. As a check, we can also obtain the energy of the black hole solution, which is the conserved quantity associated to $\xi^{a}=\partial_{t}$,
\begin{equation}
E = \oint_{\Sigma} d^{2}y \sqrt{\sigma} \xi^{a} \xi^{b} \langle\tau^{\mathrm{dual}}_{ab}\rangle = \frac{LL_{b} \mu_b}{\kappa l^{2}} ~,
\label{en2}
\end{equation}
which matches (\ref{en1}) and also the general expression obtained in \cite{Anabalon:2014fla} for these particular boundary conditions.

\subsection{The Euclidean action of the hairy AdS soliton}

For the sake of obtaining all the necessary thermodynamical quantities to investigate the phase diagram of the system, we proceed further by performing a similar analysis of the hairy AdS soliton. Let us consider the Euclidean section of (\ref{anzatsol1}), where $\beta_s$ is the periodicity of the Euclidean time, $0\leq \tau_{E}\leq \beta_{s}$,
\begin{equation}
ds^{2} = \Omega_s(x) \left[ \frac{d\tau_{E}^2}{l^{2}} + \frac{\eta_s^{2} dx^{2}}{f(x)} + f(x) d\theta^{2} + \frac{dz^{2}}{l^{2}} \right] ~.
\label{anzatsolitoneu}
\end{equation}
The contribution of the counterterm depending on the scalar field (\ref{scalarct}) is
\begin{equation}
I_{\phi}^{E} = - \frac{L L_{s}\beta_{s}}{3 \kappa l^{4}\eta_s^{3}} (1 - \nu^2) \left[ 1 + \frac{3}{4\epsilon} \right] ~,
\label{Iphi2sol}
\end{equation}
while the sum of the other terms in the action gives
\begin{equation}
I_{\mathrm{bulk}}^{E} + I_{\mathrm{surf}}^{E} + I_{\mathrm{count}}^{E} = - \frac{L\beta_{s}\Omega_s(x_h)}{4 l^{2} G_{N}} + \frac{L L_{s}\beta_{s}}{3 \kappa l^{4}\eta_s^{3}} (1 - \nu^2) \left[ 1 + \frac{3}{4\epsilon} + \frac{\alpha l^{2}}{1 - \nu^2} \right] ~.
\label{actionsAdSsol}
\end{equation}
The Euclidean action for the hairy soliton is the sum of all the terms above and, as expected, all the divergences cancel out,
\begin{equation}
I_{\mathrm{hairy\,soliton}}^{E} = - \frac{LL_{s}\alpha\beta_{s}}{6\kappa l^{2}\eta_s^{3}} = - \frac{2\pi L}{3\kappa l^{2}} \beta_s \Omega_s(x_h) ~.
\label{Isol}
\end{equation}
The free energy is $\mathcal{F} = \beta_{s}^{-1} I_{\mathrm{hairy\,soliton}}^{E} = M_{\mathrm{hairy\,soliton}}$; therefore the mass of the hairy AdS soliton is, as expected, negative,
\begin{equation}
M_{\mathrm{hairy\,soliton}} = - \frac{LL_{s}\mu_{s}}{2\kappa l^{2}} ~, \qquad \mu_{s} = \frac{\alpha}{3\eta_s^{3}} ~.
\label{ensoliton}
\end{equation}
A similar computation to the one for the hairy black hole gives the following holographic stress tensor for the hairy AdS soliton:
\begin{equation}
\langle\tau^{\mathrm{dual}}_{ab}\rangle = \frac{\mu_s}{2\kappa l^{2}} \left[ - 3 \delta_{a}^{\ \theta} \delta_{b}^{\ \theta} + \gamma_{ab} \right] ~,
\label{taudualsol}
\end{equation}
which is covariantly conserved and corresponds to the stress tensor of a conformal gas. Using the component $\tau_{tt}$ we can compute the energy of the hairy AdS soliton,
\begin{equation}
M_{\mathrm{hairy\,soliton}} = \oint_{\Sigma} d^{2}y \sqrt{\sigma} \xi^{a} \xi^{b} \langle\tau^{\mathrm{dual}}_{ab}\rangle = - \frac{LL_{s} \mu_s}{2\kappa l^{2}} ~,
\end{equation}
and check that, indeed, it matches (\ref{ensoliton}).

\section{Phase transitions}

In this section, we shall study the possible phase transitions between hairy black hole and hairy AdS soliton. Since the free energy of the planar black hole with respect to the AdS spacetime is always positive, there are no first order phase transitions of the Hawking-Page kind. However, when one of the spatial directions is made compact, there exist first order phase transitions involving the thermal AdS soliton \cite{Surya:2001vj}, the (unique) minimum energy solution satisfying those boundary conditions \cite{Horowitz:1998ha, Galloway:2001uv, Galloway:2002ai, Woolgar:2016axs}.

\subsection{First order phase transitions}

Since we have obtained all the relevant thermodynamic quantities, it is possible to investigate the phase diagram of hairy black holes. Let us summarize what we did till now: we have started with a hairy planar black hole where one direction is compactified with a fixed periodicity. As expected, for the black hole there is only one constant of integration, $\eta_b$, that plays the role of a mass parameter.\footnote{The hair is secondary, there is no constant of integration related to it, though the mass depends on the parameter $\alpha$ in the Lagrangian.} In the same class of boundary conditions, there exists also a hairy soliton solution for which the periodicity of the angular coordinate is fixed such that a conical singularity is removed.

Before comparing the free energies we should make an observation regarding the horizon radius for the coordinate system we use. The metric ans\"atze (\ref{Ansatzbh1}) and (\ref{anzatsol1}) have conformal factors contributing to the radius of the black hole, $r_b$, and the AdS soliton, $r_s$,
\begin{equation}
r_{b}^{2} = \frac{\Omega_b(x_h)}{l^{2}} ~, \qquad r_{s}^{2} = \frac{\Omega_s(x_h)}{l^{2}} ~;
\label{ratio}
\end{equation}
$x_h$, in turn, does not depend on the soliton and black hole mass parameters. To compare the solutions we should fix the boundary conditions, which means that we have to impose the same periodicities at the boundary: $\beta_b = \beta_s$ and $L_s = L_b$. The same periodicity of the Euclidean time means that both solutions have the same temperature and so the hairy AdS soliton is the thermal background. The hairy black hole free energy with respect to that of the hairy AdS soliton is
\begin{equation}
\Delta \mathcal{F} = \beta_{b}^{-1} (I_{\mathrm{hairy\,BH}}^{E} - I_{\mathrm{hairy\,soliton}}^{E}) = \frac{2\pi L}{3\kappa l^{2}} \Omega_s(x_h) \left( 1 - \frac{r_b^3}{r_s^3} \right) ~,
\label{diffree}
\end{equation}
where we used (\ref{omega}), (\ref{omegasol}) and (\ref{ratio}). As expected, the sign of the free energies difference is controlled by the ratio of the scales involved, $r_{b}/r_{s}$. Since $\Delta \mathcal{F} = 0$ occurs for $r_{b} = r_{s}$, we get a critical temperature, $T_{c} = 1/L_{s}$ from (\ref{periosoliton}) and (\ref{temperature}).

The last step of our analysis is to obtain the entropy, which is given by the black hole area, and compare our results with the 'no-hair' case worked out in \cite{Surya:2001vj}. It is easy to obtain the following useful dimensionless expression for the area and temperature ratio:
\begin{equation}
\frac{\mathcal{A}}{T l^{3}} = \frac{L}{l} \frac{\eta_s}{\eta_b} \mathcal{L}(\alpha, \nu) ~,
\label{ratiosol}
\end{equation}
where we can use the horizon equation $f(x_h) = 0$ to write down an exact expression $\mathcal{L}(\alpha, \nu)$ as a function of the parameters of the potential:
\begin{equation}
\mathcal{L}(\alpha, \nu)=\frac{16\pi^{2}}{\alpha^{2}l^{4}}\biggl{[}\frac{\nu^{2}x_h^{\nu-1}}{(x_h^{\nu}-1)^{2}}\biggr{]}^{3} ~.
\label{newL}
\end{equation}
Before analyzing in detail (\ref{ratiosol}), let us pause and compare it with the result of \cite{Surya:2001vj} that corresponds to pure gravity when the modulus is turned off. As observed in \cite{Surya:2001vj}, unlike the AdS black holes with a spherical horizon topology, the area and temperature of the flat AdS black holes are independent quantities. This implies that the thermodynamical stability depends on the temperature of the black hole, but also on its size (there is another compact coordinate in the boundary besides the Euclidean time). The critical point, where the difference of free energies change sign, corresponds to $T_{c} = 1/L_{s}$, and it matches the result we obtained above for the hairy case. The reason is that in both cases, since the dual quantum field theory is conformal, the phase transition should be controlled by the ratio of the two scales.

In pure gravity, it was found that the small hot black holes are unstable and decay into small hot solitons, whereas the large cold black holes are thermodynamically stable \cite{Surya:2001vj}. The black holes that are in equilibrium with the AdS soliton can be large and hot, but also cold and small (this is going to be clear later on, but the main point is that the temperature is also controlled by the $\alpha$ parameter).

\begin{figure}[h]
\centering
\includegraphics[scale=0.36]{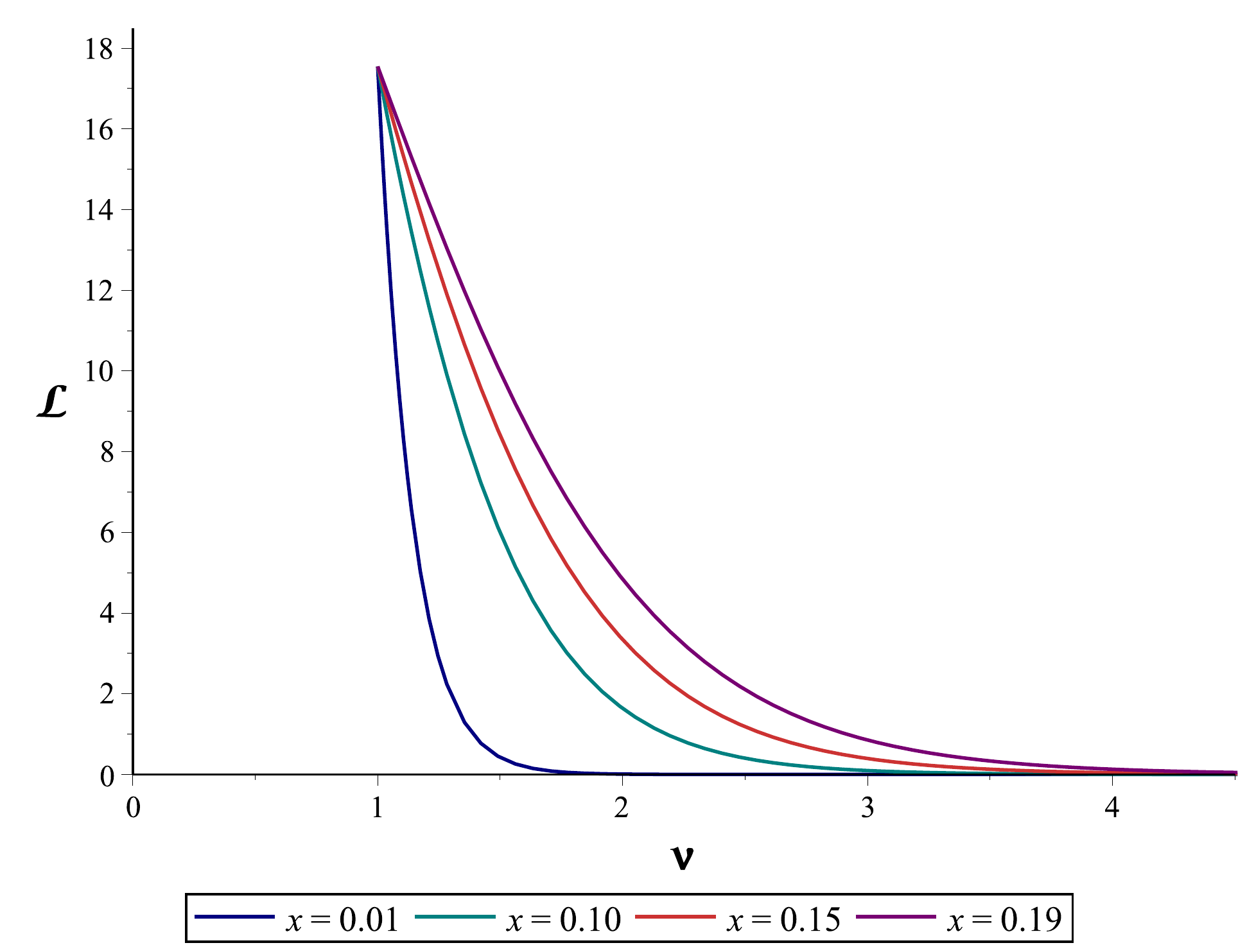} \includegraphics[scale=0.36]{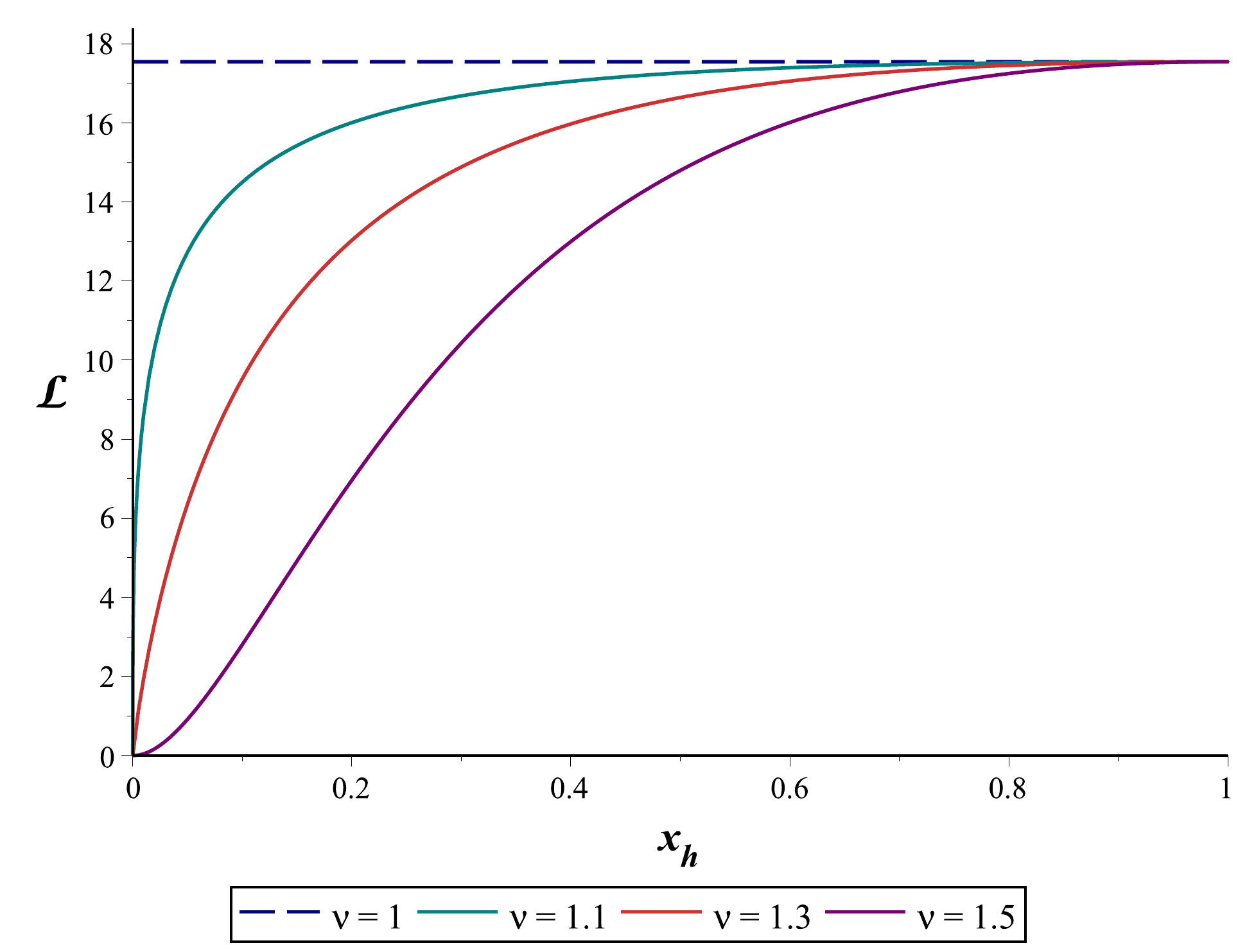}
\caption{We plot $\mathcal{L}(x_{h},\protect\nu )$ for different fixed values of $x_{h}$ (left) and $\protect\nu $ (right).}
\label{alphaLvsnux}
\end{figure}

\begin{figure}[h]
\centering-
\includegraphics[scale=0.36]{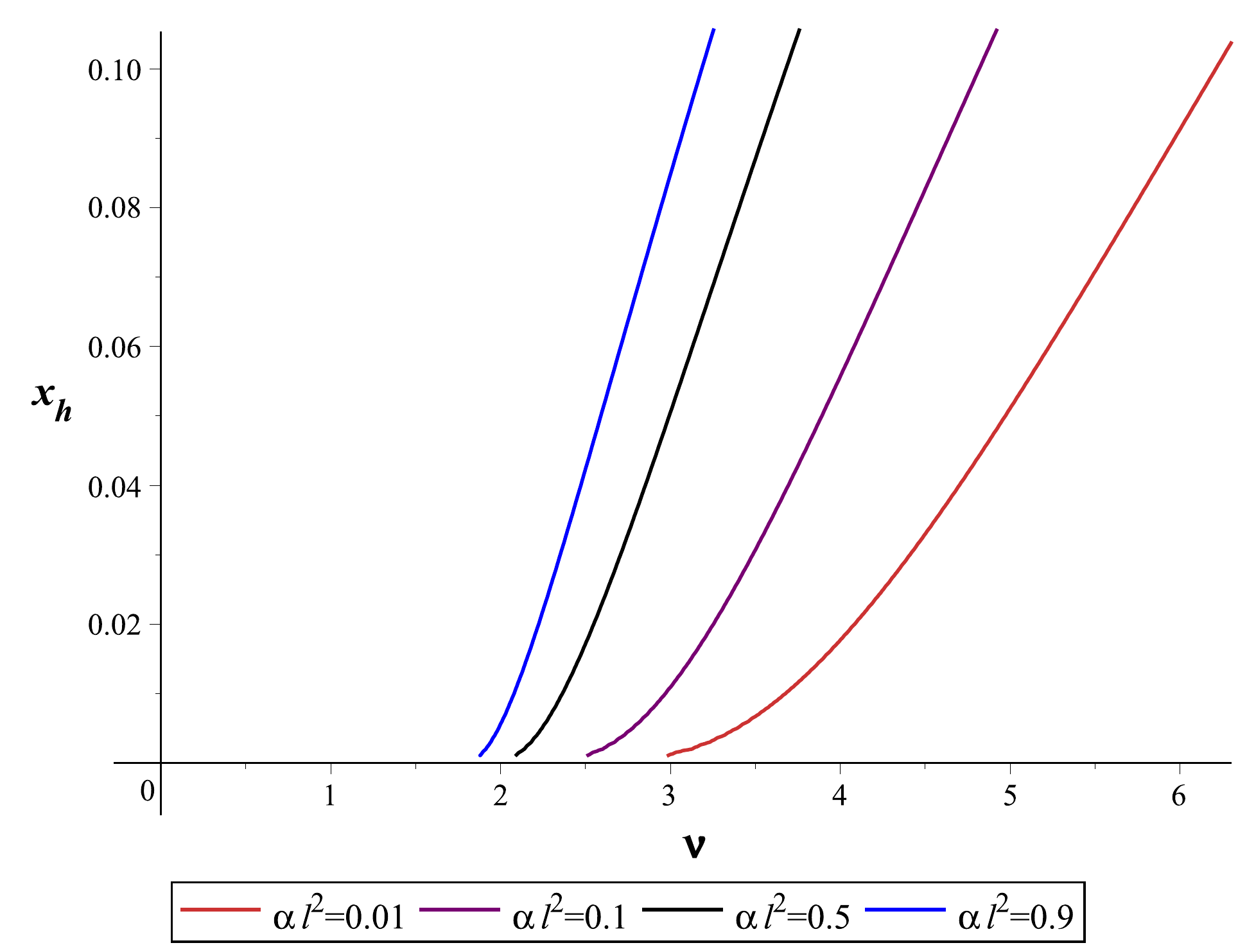} \includegraphics[scale=0.36]{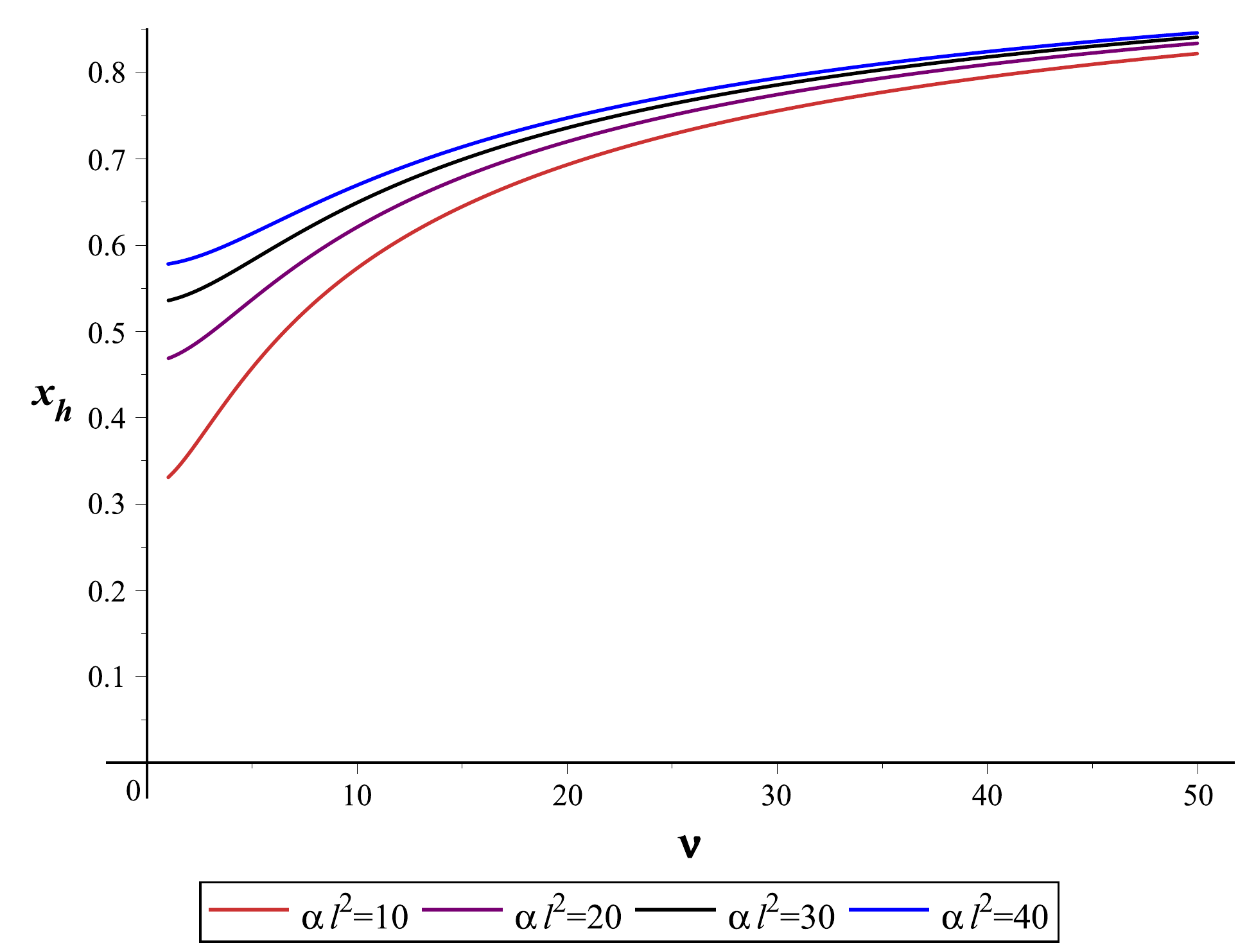}
\caption{We plot $x_{h}(\protect\nu )$ for different fixed values of $\protect\alpha l^{2}$: $\protect\alpha l^{2}<1$ (left) and $\protect\alpha l^{2}>10$ (right). We restrict ourselves to the range $x_{h}\in \lbrack 0,1]$.}
\label{alphaLvsnu}
\end{figure}

For the hairy case, let us consider $L$ of the same order as the AdS radius, $L\simeq l$. We can readily prove that $r_{b}/r_{s}=\eta _{s}/\eta _{b}$, which allows us to rewrite (\ref{ratiosol}) as
\begin{equation}
\frac{\mathcal{A}}{Tl^{3}}\simeq \mathcal{L}(\alpha ,\nu )\frac{r_{b}}{r_{s}} ~.
\label{imp}
\end{equation}
The main difference with the no-hair case is the appearance of the factor $\mathcal{L}(\alpha ,\nu )$ and so the parameters of the moduli potential are going to play an important role in describing the phases. In the remainder of this section we investigate how the presence of this factor affects the thermodynamical stability of planar hairy black holes. It is not possible to analytically solve the equation of the horizon $f(x_{h})=0$ for $x_{h}$, but we can easily obtain $\alpha =\alpha (x_{h},\nu ,l^{2})$,
\begin{equation}
\alpha l^{2}=-\biggl{[}\frac{1}{\nu ^{2}-4}-\frac{x_{h}^{2}}{\nu ^{2}} \biggl{(} 1 + \frac{x_{h}^{-\nu}}{\nu - 2} - \frac{x_{h}^{\nu}}{\nu + 2} \biggr{)} \biggr{]}^{-1} ~.
\label{newf}
\end{equation}
We can then trade the parameter $\alpha $ for $x_{h}$ and rewrite $\mathcal{L}(\alpha ,\nu )$ in (\ref{newL}) as
\begin{equation}
\mathcal{L}(x_{h},\nu )=16\pi ^{2}\biggl{[}\frac{\nu ^{2}x_{h}^{\nu -1}}{(x_{h}^{\nu }-1)^{2}}\biggr{]}^{3}\biggl{[}\frac{1}{\nu ^{2}-4}-\frac{x_{h}^{2}}{\nu ^{2}}\biggl{(}1+\frac{x_{h}^{-\nu }}{\nu -2}-\frac{x_{h}^{\nu
}}{\nu +2}\biggr{)}\biggr{]}^{2} ~;
\label{newL1}
\end{equation}
its profile is depicted in Figure \ref{alphaLvsnux}.
It is also convenient to display $x_{h}=x_{h}(\nu )$ for different values of $\alpha l^{2}$; given that we have restricted the range of the radial coordinate such that $x_{h}\in \lbrack 0,1]$, $x_{h}(\nu )$ is single-valued; see Figure \ref{alphaLvsnu}.
Let us now discuss some special cases for specific values of the parameters. which have a neat interpretation. The case $\nu = 1$, for instance, corresponds to the Schwarzschild planar black hole. This can be easily understood because in this case the scalar field vanishes and the metric components become
\begin{equation}
\Omega _{b}(x)=\frac{1}{\eta _{b}^{2}(x-1)^{2}} ~, \qquad f(x)=\frac{1}{l^{2}} + \frac{\alpha }{3}(x-1)^{3} ~.
\label{nuequalsone}
\end{equation}
Now, with the following change of coordinates
\begin{equation}
x = 1 - \frac{1}{\eta_{b}r} ~,
\end{equation}
we obtain the usual expression for the planar black hole \cite{Surya:2001vj},
\begin{equation}
\Omega _{b}(r)=r^{2}~,\qquad f(r)=\frac{1}{l^{2}}-\frac{\alpha }{3\eta_{b}^{3}r^{3}} ~.
\end{equation}
Similarly, the hairy soliton solution can be obtained by using the following coordinate transformation:
\begin{equation}
x = 1 - \frac{1}{\eta_{s}r} ~.
\end{equation}
Whenever there exists a black hole, $x_{h}$ should be a positive finite number and it is determined by the parameter $\alpha $ through the equation $\alpha l^{2}=-3/(1-x_{h})^{3}$; this can be seen in Figure \ref{alphaLvsnu} (right), where for $\nu =1$ there exists a finite $x_{h}$ that depends on the parameter $\alpha $, whenever $\alpha l^{2}>3$. In this case, we have $\mathcal{L}(x_{h},1)=16\pi ^{2}/9$, which leads to the results of \cite{Surya:2001vj} when the scalar field vanishes, see Figure \ref{alphaLvsnux}.

When $\nu \to \infty$ and $x_h$ has a finite value, $\alpha l^2 = 0$, which corresponds to $AdS$ spacetime and so this is a trivial case; see Figure \ref{alphaLvsnux} (left). Finally, let us consider $x_h \to 1$. In this case, from (\ref{newf}), we see that $\alpha l^2 \to \infty$ for any $\nu$, but interestingly we obtain a finite value for $\mathcal{L}(1,\nu) = 16\pi^2/9$; see Figure \ref{alphaLvsnux} (right). Therefore, this case is again similar to the no-hair setup discussed in \cite{Surya:2001vj}.

After this general analysis of some limiting cases of $\mathcal{L}(x_{h},\nu) $, we would like to understand the possible phase transitions when the parameter $\nu$ is fixed. This is particularly important because when, \textit{e.g.}, $\nu=4$ our theory corresponds to a (one scalar) consistent truncation of $\omega$-deformed $\mathcal{N}=8$ supergravity. We would like to point out from the beginning that the phase transitions and thermodynamic behavior change when $\nu <2$ and $\nu>2$. In what follows, we are going to discuss in detail the case $\nu<2$ and also the limiting case $\nu=2$. We postpone for the next section, where we are also going to discuss the embedding in $\omega$-deformed $\mathcal{N}=8$ supergravity, the case $\nu>2$.

Let us start with $\nu<2$. In this case, we can easily check from (\ref{newf}) that
\begin{equation}
\alpha l^{2} = \frac{(\nu^{2}-4)\,\nu^{2}}{x_{h}^{2-\nu} (\nu+2) - x_{h}^{2+\nu} (\nu-2) + x_{h}^2 (\nu^{2}-4) - \nu^{2}} ~,
\label{newf1}
\end{equation}
and so, when $x_h=0$, $\alpha l^2 = 4 - \nu^2$. This result is neatly expressed in Figure \ref{alphaLvsnu} (left), which means that for a fixed $\nu<2$, there exist black hole solutions just when $\alpha l^2 > 4 - \nu^2$.
\begin{figure}[h]
\centering
\includegraphics[scale=0.35]{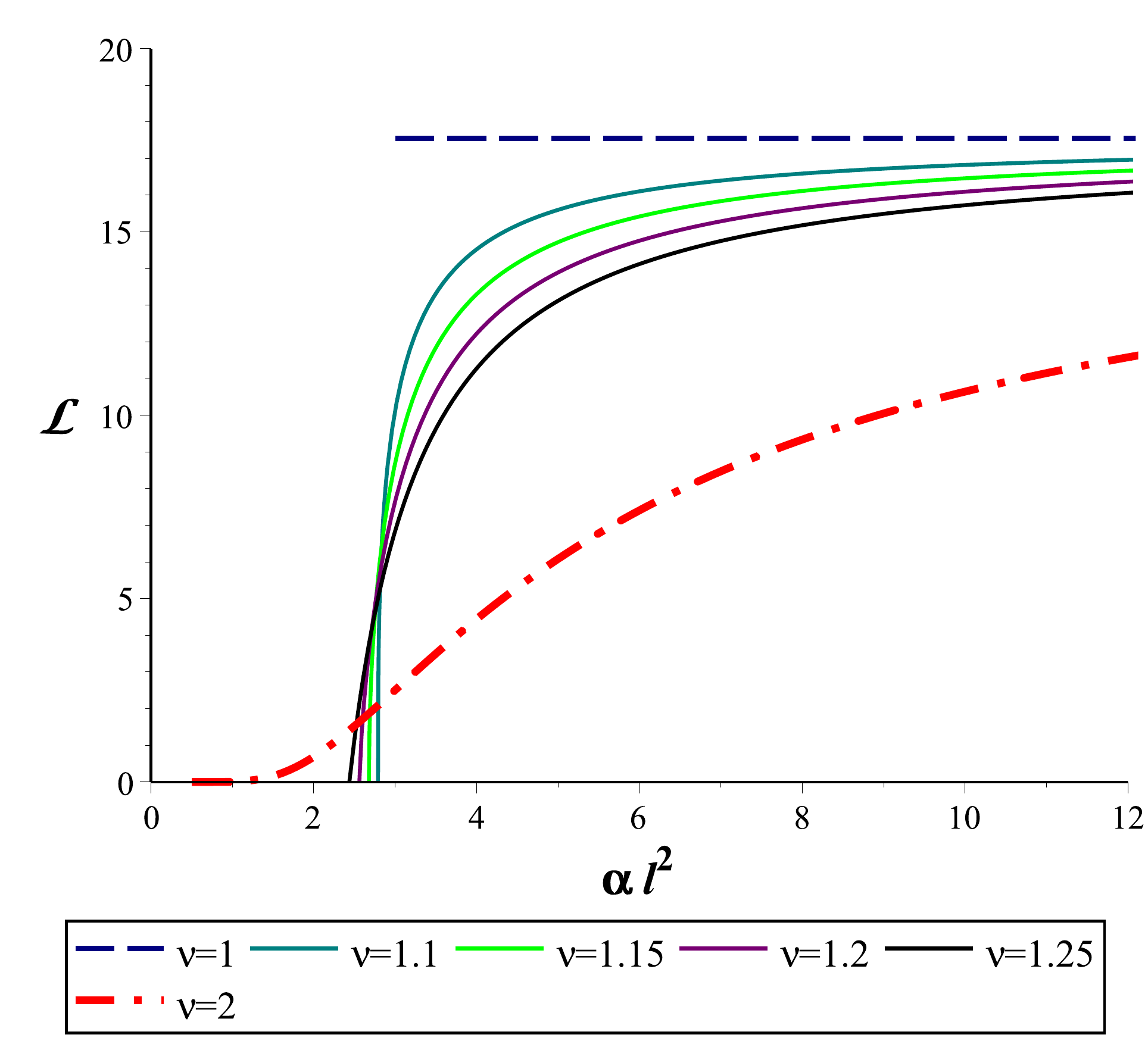} \includegraphics[scale=0.35]{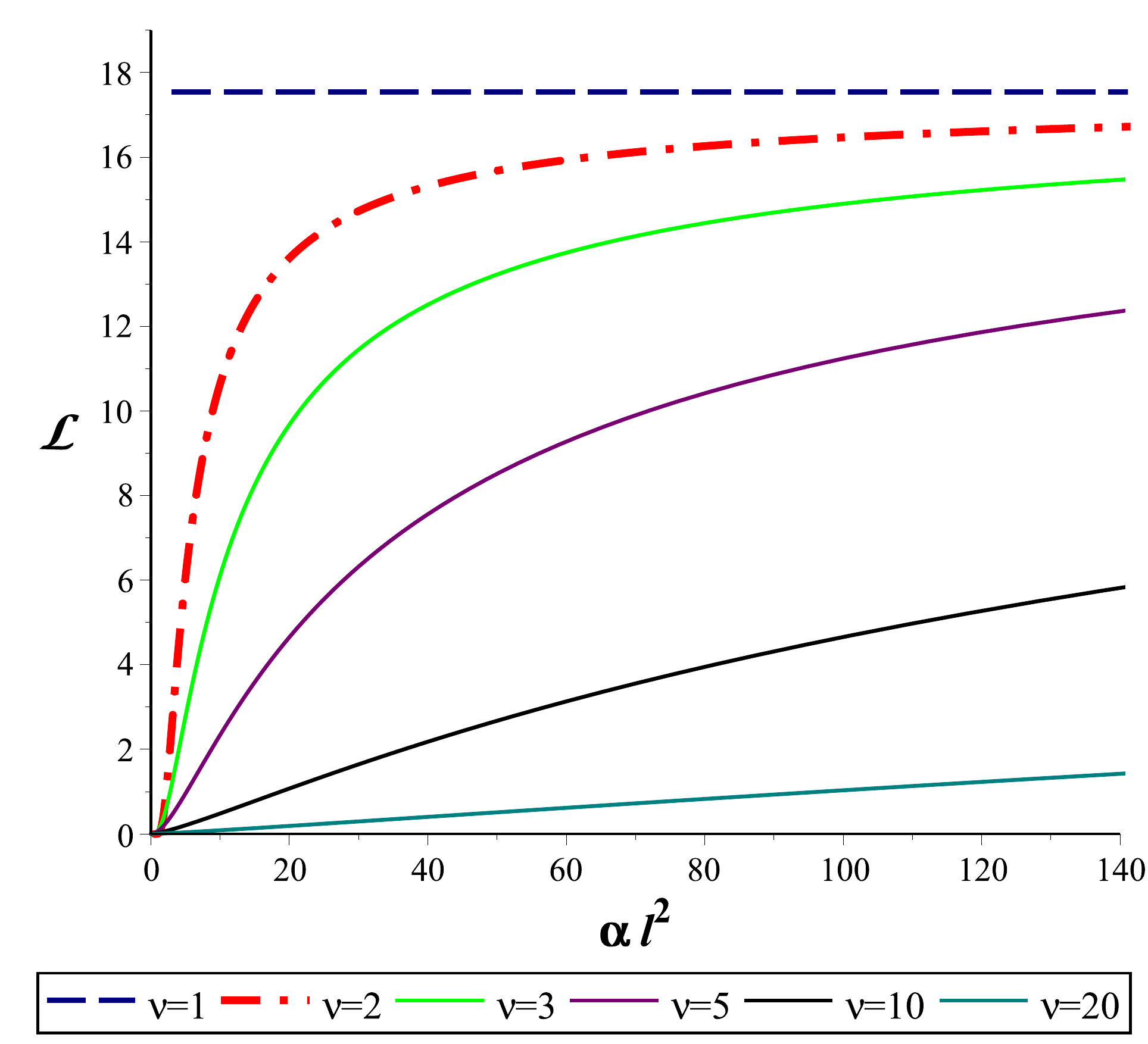}
\caption{We plot $\mathcal{L}(\protect\alpha,\protect\nu)$ for different fixed values of $\protect\nu$ by using the parametric expressions (\protect\ref{newf}) and (\protect\ref{newL1}) in terms of $x_h$, $x_{h}\in [0,1]$.}
\label{Lvsalphaa}
\end{figure}
What is important for the phase diagram is that $\mathcal{L}(x_{h} \to 0,\nu \neq 1) \to 0$, which correspond to a finite $\alpha$, as it is shown in Figure \ref{Lvsalphaa}. To understand the first order phase transitions between the hairy black hole and the thermal hairy soliton, we should consider the equations (\ref{temperature}) and (\ref{imp}). When $\mathcal{L}(x_{h},\nu) \to 0$ for a fixed $\nu \neq 1$, the parameter $\alpha$ is finite and so the the temperature of small and large black holes is not affected (\ref{temperature}). However, from (\ref{imp}) we can conclude that when $\mathcal{L}(x_{h},\nu)$ is very small the hot small black holes can also be stable. This is a drastic change if one compares with the no-hair case studied in \cite{Surya:2001vj}, where it was found that the small black holes are always unstable with respect to the AdS soliton. One explanation for this behavior could be that the self-interaction is very strong close to the horizon and it acts as a box for the small black holes; see, for instance, \cite{Nunez:1996xv}. Since for large values of $\alpha$, the factor $\mathcal{L}(x_{h},\nu)$ appearing in (\ref{imp}) is bounded from above, there is no qualitative difference between this case and the results of \cite{Surya:2001vj}.

Let us end this section with a discussion of the $\nu=2$ case. One can check that the limit $\nu \to 2$ is smooth, the scalar potential is regular and there exist hairy black hole solutions. We obtain:
\begin{equation}
\alpha l^{2}(\nu=2) = \frac{16}{4x_{h}^{2}-x_{h}^{4}-4\ln{x_{h}}-3} ~, \qquad \mathcal{L}(\alpha,2) =\frac{16\pi^{2}}{\alpha^{2}l^{4}} \biggl{[} \frac{4x_{h}}{(1 - x_h^{2})^{2}}\biggr{]}^{3} ~.
\label{alphaL2}
\end{equation}
One important difference with the cases $\nu \neq 2$ is that now $\alpha l^2 \to 0$ in the limit $x_h \to 0$, thereby small black holes can be cold. $\mathcal{L}(\alpha,2)$ vanishes when $\alpha l^2 = 0$, but it is again finite when $\alpha l^2 \to \infty$.

\subsection{Phase transitions in $\protect\omega$-deformed $\mathcal{N}=8$ supergravity}

We will see now that, due to some restrictions on the scalar potential parameters arising from the embedding of our model in $\omega $-deformed $\mathcal{N}=8$ supergravity, the first order phase transitions and thermodynamical behavior of hairy black holes are similar to those of the no-hair case. We are interested in one scalar consistent truncations of $\omega $-deformed $\mathcal{N}=8$ supergravity \cite{Anabalon:2013eaa, Tarrio:2013qga}. Let us recall that these supergravities arise from a dyonic gauging by $\hat{A}_{\mu}$ in the adjoint of $SO(8)$, in comparison to the purely electric gauging originally considered in \cite{deWit:1982bul}. That is, there is a one-parameter family of theories which result from gauging a linear combination of the electric $A_{\mu}$ and magnetic $\tilde{A}_{\mu}$ potentials, $\hat{A}_{\mu} = \cos \omega \,A_{\mu}+\sin \omega \,\tilde{A}_{\mu}$, physically distinct theories belonging to the range $\omega \in \lbrack 0,\frac{\pi }{8}]$ \cite{DallAgata:2012mfj}; of course, $\omega =0$ corresponds to the theory developed in \cite{deWit:1982bul}.

The only case within maximal supergravity where we have been able to find analytic solutions is the potential that breaks the isometries of the $S^{7}$ to $SO(5)\times SO(3)$, whose action reads
\begin{equation}
I[g_{\mu\nu}, \phi] = \int_{\mathcal{M}}d^{4}x \sqrt{-g} \left[ \frac{R}{2\kappa} - \frac{1}{2}(\partial\phi)^{2} - V_\omega(\phi) \right] ~,
\end{equation}
where the potential is $V_\omega(\phi) = \cos^2\omega\, Q(\phi) + \sin^2\omega\, Q(-\phi)$, with
\begin{equation*}
Q(\phi) = - \frac{3}{16l^{2}} \left[ e^{-\phi \sqrt{\frac{10}{3}}} + 5e^{4\phi \sqrt{\frac{2}{15}}} + 10e^{-\phi \sqrt{\frac{2}{15}}} \right] ~,
\end{equation*}
which correspond to our potential (\ref{potential}) with $\nu =4$ and $\alpha = -\frac{12 \sin^2\omega}{l^{2}}$. As stated earlier, the action is invariant under the combined transformation $\phi \to -\phi$ and $\alpha \to - \alpha - \frac{12}{l^{2}}$; {\it i.e.}, $\sin^2\omega \to \cos^2\omega$. The black hole solution follows the same pattern described above, namely:
\begin{equation}
ds^{2} = \Omega_b(x) \left[ -f(x) dt^{2} + \frac{\eta_b^{2} dx^{2}}{f(x)} + \frac{dy^{2}}{l^{2}} + \frac{dz^{2}}{l^{2}} \right] ~,
\end{equation}
where
\begin{equation}
\Omega_b(x) = \frac{1 6x^3}{\eta_b^{2}(x^4 - 1)^2} ~, \qquad f(x) = \frac{1}{l^{2}} - \frac{1}{8 l^{2}} \frac{(x^2 + 3) (x^2 - 1)^3}{x^2} \sin^2\omega ~, 
\end{equation}
and a scalar field profile:
\begin{equation}
\phi = \sqrt{\frac{15}{2}} \ln x ~.
\end{equation}
There is another solution \cite{Anabalon:2017yhv}, which can be obtained from the previous one by $\phi \to -\phi$ and $\sin^2\omega \to \cos^2\omega$. Both exist only when $x\in \lbrack 1,\infty )$ \cite{Anabalon:2013eaa}.

We check how $\mathcal{L}(\alpha ,\nu )$ varies as a function of the
parameter $\alpha $, see Figure \ref{Lvsalphaa} (right). It is clear that $%
\mathcal{L}(\alpha \rightarrow 0,\nu )\rightarrow 0$ therefore small black
holes can be stable for $\alpha l^{2}\rightarrow 0$, whereas $\mathcal{L}%
(\alpha \rightarrow \infty ,\nu )\rightarrow 16\pi ^{2}/9$, which means that
big black holes can be unstable with respect to the AdS soliton when $\alpha 
$ is sufficiently large. Only in these extreme regimes the thermodynamical
behavior can differ from the no-hair case. The case with $\alpha
l^{2}\rightarrow 0$ is the only one relevant for gauged
supergravity.

\subsection{M-theory embedding}

It turns out that the theory we have been dealing with corresponds, whenever $\alpha = 0$, to a sphere consistent reduction of 11-dimensional supergravity. The scalar fields in maximal supergravity parameterize the coset $SL(8,\mathbb{R})/SO(8)$, and the local $SO(8)$ transformations can be used to diagonalize the scalar potential such that we are led to consider the following action \cite{Cvetic:1999xx, Cvetic:2000eb},
\begin{equation}
I[g_{\mu\nu},\vec\phi] = \int_{\mathcal{M}} d^{4}x \sqrt{-g} \left[ \frac{R}{2\kappa} - \frac12 (\partial\vec\phi)^2 - V(\vec\phi) \right] ~,
\label{actionCv}
\end{equation}
whose potential is given by
\begin{equation}
V(\vec\phi) = - \frac{g^2}{4} \left[ \left( \sum_{i=1}^8 X_i \right)^2 - 2 \sum_{i=1}^8 X_i^2 \right] ~,
\end{equation}
where
\begin{equation}
X_i = \exp{\left( - \frac12 \vec b_i \cdot \vec\phi\right)} ~, \qquad \prod_{i=1}^8 X_i = 1 ~,
\end{equation}
$\vec b_i$ being the weight vectors of the fundamental representation of $SL(8,\mathbb{R})$. If we consider now a single scalar field reduction preserving SO($p$) $\times$ SO($8-p$),
\begin{equation}
X_1 = \cdots = X_p = X := e^{\frac{1}{\sqrt{2}} \sigma \phi} ~, \qquad X_{p+1} = \cdots = X_8 = Y := e^{- \frac{1}{\sqrt{2} \sigma} \phi} ~,
\end{equation}
with
\begin{equation}
\sigma = \sqrt{\frac{\nu-1}{\nu+1}} ~, \qquad {\rm and} \qquad p = \frac{4(\nu+1)}{\nu} ~,
\end{equation}
the previous action (\ref{actionCv}) reduces to the one we are studying in this paper. The action is invariant under $\sigma \to 1/\sigma$, $\phi \to -\phi$ and $p \to 8-p$. The case $\nu = 4$ is exactly the one discussed in the previous subsection, thereby we can uplift those results to 11-dimensional supergravity by means of the formulas presented in \cite{Cvetic:1999xx, Cvetic:2000eb}.

\section{Holography of hairy planar solutions}
In this section we are going to obtain the RG flow for the hairy planar black hole solution and check that when the hair parameter is $\nu=\pm 1$ and the solution becomes Schwarzschild-AdS, the flow is trivial. 

Let us start by computing the c-function that accounts for the decreasing number of degrees of freedom along the corresponding RG-flow. This function was constructed in various situations \cite{Skenderis:1999mm, deBoer:2000cz, Acena:2012mr} and here we apply the analysis to our cases of interest. For a general ansatz
\begin{equation}
ds^{2}=-h(r)^{2}dt^{2}+\frac{dr^{2}}{p(r)^{2}} + \frac{c(r)^{2}}{l^2} \left( dy^{2} + dz^{2} \right) ~,
\label{antz}
\end{equation}
a geometrical construction of the c-function was proposed in \cite{Freedman:1999gp}. Let us know consider the following combination of the $tt$- and $rr$-components of the Einstein and stress-energy tensors: 
\begin{equation*}
G_{t}^{\ t} - G_{r}^{\ r} = \frac{2p}{hc}(hpc^{\prime\prime} + h p^{\prime}c^{\prime} - p h^{\prime}c^{\prime}) ~, \qquad T_{t}^{t}-T_{r}^{r}=-\phi^{\prime 2}p^{2} ~,
\end{equation*}
which implies that
\begin{equation}
\kappa\phi^{\prime 2} = \frac{2c^{\prime}}{c} \bl{[}\ln{\bl{(}\frac{h}{pc^{\prime}}\br{)}}\br{]}^{\prime} ~.
\end{equation}
The null energy condition states that for any null vector $n^{\alpha}n_{\alpha}=0$, the stress tensor should be so that $T_{\alpha\beta}n^{\alpha}n^{\beta}\geq 0$, which for a gravity theory with a scalar field becomes $\forall p$, $\rho+p\sim \phi^{'2}\geq 0$. Using the condition $\frac{c^{'}}{c}\geq 0$, a straightforward consequence of the null energy condition is
\begin{equation}
\pa_{r}\bl{[}\ln{\bl{(}\frac{h}{pc^{'}}\br{)}^{2}}\br{]} \geq 0 ~.
\label{cfunc}
\end{equation}
Since for an RG flow one integrates out degrees of freedom from UV towards IR, let us consider that there exists a function, $\mathcal{C}(r)\geq 0$, which is monotonically increasing from the bulk to the boundary $\mathcal{C}^{\prime}(r)\geq 0$, thereby $[\ln{\mathcal{C}(r)}]^{\prime} \geq 0$. We can identify the c-function as
\begin{equation}
\mathcal{C}(r) = \mathcal{C}_{0}\bl{(}\frac{h}{pc^{\prime}}\br{)}^{2} ~,
\label{cfunction}
\end{equation}
where $C_0$ is a constant that can be determined, for example, from the two-point function of the stress tensor. Now, for our general ansatz (\ref{Ansatzbh1}), we can make the identification $h^{2} \to \Omega_b f$, $c^{2} \to \Omega_b$ and $\eta_b\,p\,\pa_{r} \to = (f/\Omega_b)^{1/2}\,\pa_{x}$, which leads to the c-function:
\begin{equation}
\mathcal{C}(x) = \mathcal{C}_{0} \bl{(} \frac{2 \eta_b \Omega_b^{3/2}}{\Omega_b^{\prime}} \br{)}^{2} = \mathcal{C}_{0} \bl{[} \frac{x^{\frac{\nu+1}{2}}}{\nu-1+x^{\nu}(\nu+1)} \br{]}^2 ~.
\label{cfuncionBH}
\end{equation}
We can easily check that for $\nu=\pm 1$, we obtain $\mathcal{C}(x) = \mathcal{C}_{0}/4$ and so the flow is indeed trivial.

\section{Conclusions and further directions}

We have investigated the phase diagram of a general class of hairy black holes and their thermodynamic properties. In some particular cases, we provide the embedding in supergravity. Witten has argued \cite{Witten:1998zw} that the Hawking-Page first order phase transition \cite{Hawking} (when the horizon topology is spherical) corresponds to a confinement/deconfinement transition of the dual strongly coupling field theory. Similarly, in the nice work \cite{Surya:2001vj}, it was shown that there also exist first order phase transitions between planar black holes and the AdS soliton. We construct the hairy AdS soliton and show that, even in the presence of the scalar field and its self-interaction, there exist first order phase transitions. 

\begin{figure}[h]
	\centering
	\includegraphics[scale=0.45]{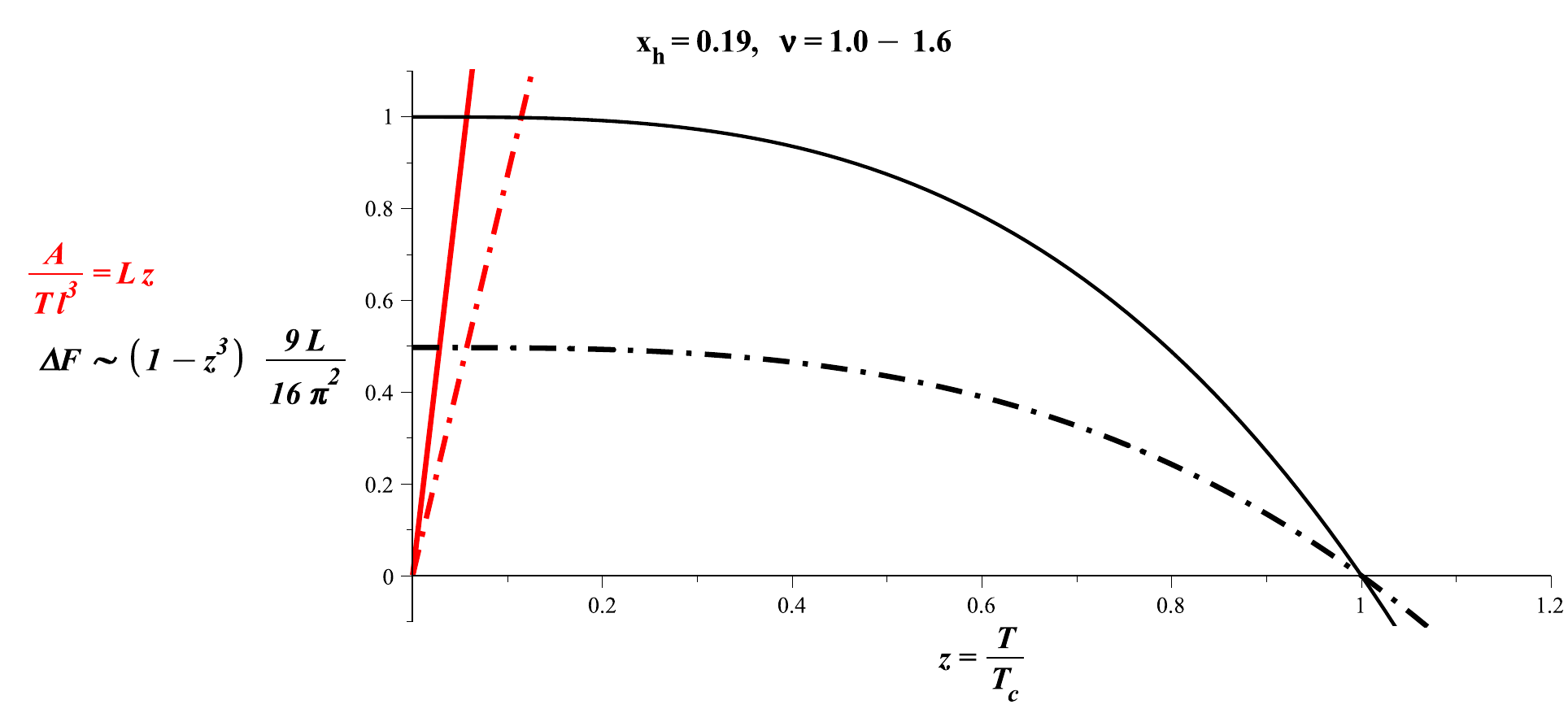} 
	\caption{ The continuous lines  represent the case $\nu=1$. We observe that, by increasing $\nu$, the free energy is decreasing. When $z<1$, in the hairy case we have $\mathcal{A}<Tl^{3}$ for $z<\frac{1}{\mathcal{L}}{\vert}_{x_{h}=0.19,\nu=1.6}\approx 0.11$. This means that even if the small hairy black holes can be colder than the planar black hole without hair for which $z<0.06$, the hairy AdS soliton is still the preferred phase.}
\label{freee}	
\end{figure}

We can summarize our results by rewriting the free energy and horizon area in a more compact form
\begin{equation}
\Delta\mathcal{F} \sim (1-z^{3})\frac{9\mathcal{L}(x_{h},\nu)}{16\pi^{2}}, \qquad \frac{\mathcal{A}}{Tl^{3}}=z\mathcal{L}(x_{h},\nu)
\end{equation}
where $z=T/T_c=TL_s$. For the usual planar black hole when the hair is turned off, $\nu=\pm 1$, we obtain $\mathcal{L}=16\pi^{2}/9\approx 17.5$ and so
\begin{equation}
\Delta\mathcal{F}\sim (1-z^{3}), \qquad \frac{\mathcal{A}}{Tl^{3}}=z~\frac{16\pi^2}{9}
\end{equation}
Even if in both cases the free energy vanishes at a critical temperature, which in the hairy case is not affected by the parameter $\nu$, we observe that the free energy depends on it. As we have emphasized in Section 3, the free energy can be made arbitrarily small by changing $\nu$ --- we do not repeat that discussion here, but we present in Fig. \ref{freee} the behaviour of the free energy and the ratio of area over temperature for two different values of the hair parameter. 

It will be interesting to extend our analysis to the charged hairy black hole solutions \cite{Anabalon:2013sra, Lu:2013ura}. Another interesting direction is to try constructing exact solutions with more general mixed boundary conditions for the scalar field. It is worth noting that, in \cite{Caldarelli:2016nni}, new exact black brane solutions with mixed boundary conditions for the scalar field were obtained. The moduli potential used in \cite{Caldarelli:2016nni} is a particular case of the general potential proposed in \cite{Anabalon:2013sra}. There exist more general boundary conditions for which one should introduce logarithmic terms in the fall off of the fields and it was proven  in \cite{Caldarelli:2016nni} that, in this case, the regularization procedure should be supplemented with extra counterterms similar to the ones in \cite{Anabalon:2015xvl}.

\section*{Acknowledgements}
%
Research of AA is supported in part by Fondecyt Grants 1141073 and 1170279. The work of DA is supported by the Fondecyt
Grants 1161418 and 1171466.
DC is supported by Fondecyt Postdoc Grant 3180185. He wishes to thank Universidad Nacional de San Antonio Abad del Cusco, during some stages of this research which is supported by the Peruvian Government through Financing Program of CONCYTEC.
The work of J.D.E. is supported by the Ministry of Science grant FPA2017-84436-P, Xunta de Galicia ED431C 2017/07, FEDER, and the Mar\'\i a de Maeztu Unit of Excellence MDM-2016-0692. He wishes to thank Pontificia Universidad Cat\'olica de Valpara\'\i so and Universidad Adolfo Ib\'a\~nez for hospitality, during the visit funded by CONICYT MEC 80150093, at the initial stages of the article.

\appendix

\end{document}